\documentstyle[aps,epsf,epsfig,rotate,prl]{revtex}
\newcommand{\be}{\begin{equation}}
\newcommand{\ee}{\end{equation}}
\newcommand{\ba}{\begin{eqnarray}}
\newcommand{\ea}{\end{eqnarray}}

\draft        
\begin{document}             
\title{Dimer state of spin-$1$ Bosons in an optical lattice}
\author{S. K. Yip}
\address{Institute of Physics, Academia Sinica, Nankang, Taipei 11529, Taiwan}
\maketitle

\begin{abstract}

In this paper we consider spin-$1$ Bosons, such as $^{23}$Na,
trapped in an optical lattice, in the regime of one particle per site.
We argue that the ground state is expected
to be the dimer phase in one, two, or three dimensions,
thus realizing a state that has so far been studied only theoretically.


\end{abstract}

Tremendous progress has been made recently on trapping 
and cooling atoms.  Greiner et al \cite{Greiner02} has succeeded
in observing the Bosonic Mott-Hubbard transition.  In the 
experiment, the Bosons involved are $^{87}$Rb atoms and the lattice
potential is provided by the standing waves of three orthogonal
laser beams.  For weak lattice potentials, the system remains
a superfluid (Bose-Einstein condensate).  When the lattice
potential is increased beyond a critical strength, the tunneling
rate for atoms between the different wells becomes weak compared
with the repulsion for two atoms residing in the same well,
the system enters into the Mott regime.  In this case
the number of atoms in each well is essentially fixed 
at an integer value (here, one).
Phase coherence and hence superfluidity is lost.
\cite{Fisher89,Jaksch98}

The $^{87}$Rb atoms in the experiment of Greiner et al \cite{Greiner02}
were hyperfine-spin-polarized.
 In this paper, we discuss the interesting
physics that can be realized if $^{23}$Na atoms are employed
instead of $^{87}$Rb and if the atoms are not polarized.
Bose-Einstein condensation of unpolarized $^{23}$Na has already
been achieved by the MIT group \cite{Stenger99}. The
$^{23}$Na atoms
have hyperfine spin (hereafter simply as "spin") $1$
 in their lower energy manifold, and
the interaction among them is antiferromagnetic \cite{Ho98,Stenger99} 
(in contrast to $^{87}$Rb, where it is ferromagnetic).
  We shall confine
ourselves to the Mott regime where there is essentially one atom
per cell, and discuss the arrangement of the hyperfine spin
states (or, more precisely, the projections) for the atoms in this lattice.
We shall argue that the ground state is expected
to be the dimer phase in one, two, or three dimensions.
This is in contrast
with spin-$1$ electronic systems where the effective Hamiltonian
is the antiferromagnetic Heisenberg model.  There in one dimension,
the system would be in the Haldane phase, whereas 
long range Ne\'el order is expected to be present in
two and three dimensions. \cite{Auerbach94}

Consider then a (cubic) lattice 
formed by three orthogonal laser beams,
with effective potential for the $^{23}$Na atoms of the form
$V(x,y,z) = V_x {\rm sin}^2 (kx) + V_y {\rm sin}^2 (ky) 
               + V_z {\rm sin}^2 (kz)$
where $k \equiv { 2 \pi \over \lambda}$ 
is the wave-vector of the lasers.  The strength
$V_{x,y,z}$ of the sinusoidal potentials are proportional
to the intensities of the laser beams and can be adjusted 
separately. 
As in Ref \cite{Jaksch98} we assume that only one orbital state is
involved for each well.  
The effective Hamiltonian for our system can be
written in a Bose-Hubbard form \cite{Jaksch98}
 (generalized to Bosons with spin).
 The hopping matrix
elements, $t_{x,y,z}$ are dependent on the directions 
of hopping.  The order of magnitude of $t_x$ is given
by $ |t_x| \sim (E_R V_x)^{1/2} e^{ - 2 ( V_x / E_R )^{1/2}}$
where $E_R \equiv \hbar^2 k^2 / 2 m $ and $m$ is the mass of
the atoms.  Though we are considering one particle per well,
the system can exist in excited states 
where the wells are multiplely occupied. 
We shall confine ourselves to the regime where these
energies are large compared with the hopping energies
$|t_{x,y,z}|$. Thus we need only consider those excited states with at 
most two particles per well. 
 The extra energy is described by
the Hubbard repulsion $U_S$ which depends on the total spin $S$
of the two particles involved.  $U_S \sim ( E_R V_x V_y V_z)^{1/4} k a_S$
where $a_S$ is the scattering length in total spin $S = 0, 2$ channel.
Excited states with $S = 1$ are
 not allowed due to the identity of the Bosons with
one orbital state per well.

  We next construct, in the standard manner,
 the effective Hamiltonian for the spins in the subspace
of exactly one particle per well, assuming $|t| << U_{0,2}$
($ << $ the excited energies $ \sim (E_R V_{x,y,z})^{1/2} $ 
of higher orbital states in the wells:  this latter inequality
is typically satisfied since $a_S << \lambda$ \cite{Jaksch98}).
Let us first consider two sites, labelled by $1$ and $2$.
For $t \to 0$ the energy is independent of the spin configurations.
For finite but small $t$, we can perform perturbation in $t$.
It is convenient to classify the states according to the total
spin $S_{\rm tot}$ of the two sites.
To second order in $t$, the energy is lowered due to hopping
by the amount $ - 4 t^2 / U_{0,2}$ for total spin 
$S_{\rm tot} = 0, 2$ \cite{com1} but zero for total spin $1$.
With ${\bf S}_{1,2}$ the spin-$1$ operators for sites $1,2$ and 
using $ {\bf S}_1 \cdot {\bf S}_2 = -2, -1, 1$ for
$S_{\rm tot} = 0, 1, 2$ respectively, we find that
the effective Hamiltonian can be written as
\be
\hat H_{12} = \epsilon + J ( {\bf S}_1 \cdot {\bf S}_2 )
    \  +  \ K ( {\bf S}_1 \cdot {\bf S}_2 )^2
\label{H12}
\ee
where
$J = - { 2 t^2 \over U_2}$, $K = - { 2 \over 3} { t^2 \over U_2}
        - { 4 \over 3} { t^2 \over U_0 } $
and $\epsilon = J - K $.  
For $^{23}$Na atoms, $U_2 > U_0 > 0$ \cite{Ho98,Stenger99,comU},
hence $ K < J < 0$. 
 As a matter of fact, since $a_2 \approx 52 a_B$ and
$a_0 \approx 46 a_B$ (here $a_B = $ Bohr radius),
$ (U_2 - U_0)  / U_0  = (a_2 - a_0) / a_0  <<  1$.  Hence
$|J| \sim |K|$ and 
$|J-K| << |K|$ or $|J|$.
[For $^{87}$Rb, $U_0 > U_2 > 0$, in that case $ J < K < 0$ ].

Thus the Hamiltonian (\ref{H12}) differs significantly from the Heisenberg
one \cite{Zieger73} familiar in ionic crystals.
There, although the total spin of an ion can be larger than
$1/2$ and hence 
$ ({\bf S}_1 \cdot {\bf S}_2 )$ and $({\bf S}_1 \cdot {\bf S}_2 )^2$
are independent operators, the Hamiltonian can, to a good approximation
(ignoring spin-orbit interactions etc), be written simply as
$\hat H_{Hei} = J ( {\bf S}_1 \cdot {\bf S}_2 )$.  
For perturbation to second order in the hopping of electrons between the two
ions, there cannot be terms in $ {\bf S}_1 \cdot {\bf S}_2 $
with powers higher than $1$ in the effective Hamiltonian,
since two hoppings of electrons with spin $1/2$
 can only change the $z$ component
of the spin of an ion by at most $1$, whereas, 
e.g., $ ({\bf S}_1 \cdot {\bf S}_2 )^2 $ consists of terms
which can change that component by $\pm 2$.
Higher order perturbation in hopping can give rise to
terms higher powers in $ ({\bf S}_1 \cdot {\bf S}_2 )$
but only with much smaller coefficients when hopping
is small compared with the Hubbard repulsion.

Another peculiar fact about Hamiltonian (\ref{H12}) is also apparant.
If the spins were classical vectors, $K < J < 0$ would
require that the two spin vectors be {\it parrallel} in the lowest
energy state.  However, since $U_2 > U_0 > 0$,
$E_{S=0} < E_{S=2}$ and hence the spins on the two sites
actually prefer to be {\em anti}-ferromagnetically
correlated.  The strong quantum mechanical nature
of the spins will be of significance below.

Let us now begin with the case where
$V_{x} << V_{y,z}$, so that one can ignore couplings along
$y$ and $z$ directions.  We thus have a collection of one-dimensional
spin-$1$ chains.  Ignoring the $\epsilon$ term in eq (\ref{H12})
not of relevance below, our effective Hamiltonian is thus (for one chain)
\ba
\hat H &=& \sum_l \ \left[ 
   \ J \ ( {\bf S}_{l} \cdot {\bf S}_{l+1} )
  \  + \ K \ ( {\bf S}_{l} \cdot {\bf S}_{l+1} )^2 \ \right]
\label{H} \\
&\equiv& \sqrt{J^2 + K^2} \  
\sum_l \ \left[ \ {\rm cos} \gamma \ ( {\bf S}_{l} \cdot {\bf S}_{l+1} ) \
   + \ {\rm sin} \gamma \ ( {\bf S}_{l} \cdot {\bf S}_{l+1} )^2 \ 
   \right]
\label{H1D}
\ea
where the sum is over the site labels $l$ and
the second relation defines $\gamma$. For $ K < J < 0$  
$ - { 3 \pi \over 4} < \gamma < - {\pi \over 2}$.
If $|U_2 - U_0| << U_0 {< \atop \sim} U_2$ as for $^{23}$Na,
$\gamma \to - { 3 \pi \over 4}$, whereas if
$ U_2  >> U_0 > 0 $, $\gamma \to - {\pi \over 2}$. 
[For $ J < K < 0$
as in $^{87}$Rb,  
$  - {\pi} < \gamma < - { 3 \pi \over 4} $].

Hamiltonian of the form in (\ref{H1D}) has been studied before, with
most efforts in the region where the ground state is
expected to be in the Haldane phase ( $ - {\pi \over 4} < \gamma
 < {\pi \over 4}$).\cite{review}
  There has been much fewer studies in the range
of $\gamma$ of relevance here.  Though it is agreed that
for $\gamma$ near $ - {\pi \over 2}$ the ground state should be
in a dimer state \cite{Parkinson88}, results 
for other $\gamma$'s have been somewhat controversial,
especially 
near $ - { 3 \pi \over 4}$.  In particular Chubukov \cite{Chubukov91}
suggested that there is a critical $\gamma_c$ 
 ( $ - {3 \pi \over 4} < \gamma_c< -{\pi \over 2}$) such that
the dimer state is unstable for $\gamma < \gamma_c$, where
instead the nematic phase should exist.
A later study \cite{Fath93} claims otherwise.
Recently Demler and Zhou \cite{Demler02} considered possible ground states
for spin-$1$ Bosons
in optical lattice as in the present paper, did not
include the dimer state in their discussion.

\begin{figure}[h]
\epsfig{figure=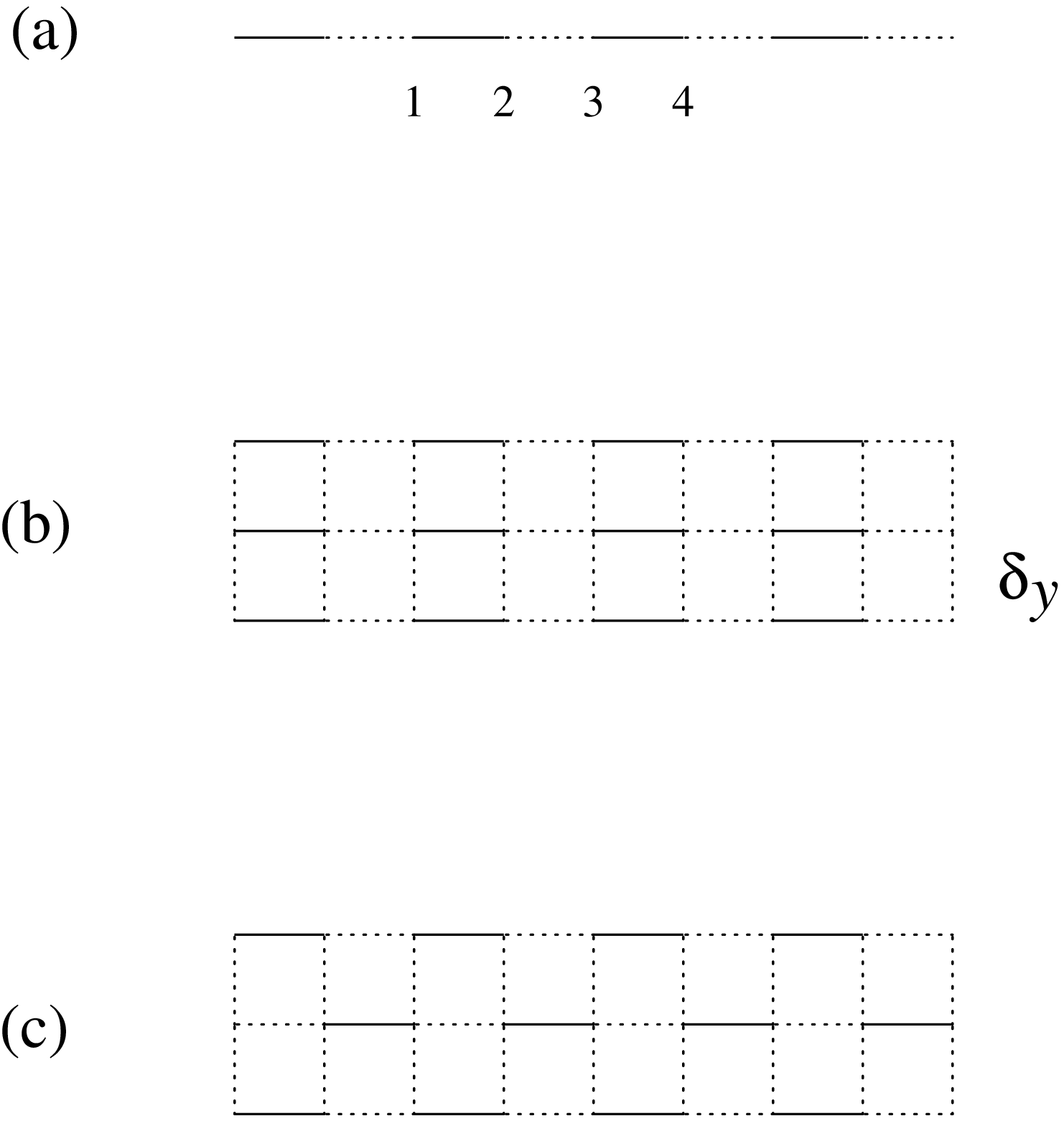,width=6in,angle=-90}
\vskip 0.4 cm
\begin{minipage}{0.95\textwidth}
\caption[]{ Dimers.  Thick lines represent
the pairs such as eq (\ref{zeta}). 
 }
\label{fig:dimers}
\end{minipage}
\end{figure}

We would like to first re-address this issue by variational ansatz.
We shall denote the three possible
spin projection states at a given site by $|+>$,$|0>$ and $|->$,
and write the wavefunctions by specifying these states at each site.
The nematic phase is given by 
\be
\Psi_{Nem} = | \ ... \ 0 0 0 0 \ ... \ >
\label{Nem}
\ee
This state is in direct analogy with the corresponding "polar" phase
\cite{Ho98} in the bulk.  This state has $ < {\bf S}_l > = 0$ and
$ < S_{l,x}^2 > = < S_{l,y}^2 > \ne < S_{l,z}^2>$.  \cite{LRNO}
For the dimer state, the usually employed ansatz is 
\cite{review,Chubukov91}
\be
\Psi_{Dimer} = \ ... \ \Psi_{12} \ \Psi_{34} \ ...
\label{Dimer}
\ee
where
\be
\Psi_{12} = { 1 \over \sqrt{3} }  ( | + - > + | - + > - | 0 0 > )_{12}
\label{Psi0}
\ee
is a singlet ( $ S_{\rm tot} = 0$) pair formed by sites $1$ and $2$
(the subscripts label the sites).
This state is shown schematically in Fig \ref{fig:dimers} (a).
(There is another state degenerate with (\ref{Dimer}) with
all pairs shifted by one lattice site. )
In the
{\it non-interacting} spin-wave approximation, the
the $S = 2$ modes become unstable at $\gamma_c = - \pi + {\rm tan}^{-1}
{ 9 \over 5} \approx - 0.66 \pi$.  Chubukov \cite{Chubukov91}
 then concludes that for
$ - { 3 \pi \over 4} < \gamma < \gamma_c <  - { \pi \over 2}$,
(with $\gamma_c$ possibly renormalized), the dimer state
is unstable, and further speculates that the correct ground
state should be the nematic state.

 Here we shall revisit this question of stability
by studying an improved variational ansatz.
 We shall write again eq (\ref{Dimer}) but with 
the pair wavefunction given by
\be
\Psi_{12} = { 1 \over \sqrt{2 + |\zeta|^2} }  \ 
 \left( \ | + - > + | - + > - \ \zeta \ | 0 0 > \right)_{12}
\label{zeta}
\ee
etc., with $\zeta$ our variational parameter.
  If $\zeta = 1$, our ansatz reduces to that in eq. (\ref{Psi0}).
Notice that if $|\zeta| \to \infty$, then 
$\Psi_{Dimer} \to \Psi_{Nem}$.  It is simple to evaluate
the expectation values of $\hat H$.  We find, with $\zeta = |\zeta| e^{i \phi}$,
\be
< \hat H_{12} > = { 2 \over 2 + | \zeta |^2 }
 \left\{ - ( 1 + 2 |\zeta| {\rm cos} \phi ) {\rm cos} \gamma
  +
  ( 3 +  2 |\zeta| {\rm cos} \phi + | \zeta  | ^2 ) {\rm sin} \gamma
  \right\}
\label{h12}
\ee
for the bond (the part of $\hat H$) between sites $1$ and $2$, and
\be
< \hat H_{23} > = { 2 \over [2 + | \zeta |^2]^2 }
\left\{  ( 3 +  2 |\zeta|^2 + | \zeta  | ^4 ) {\rm sin} \gamma
  \right\}
\label{h23}
\ee
for the bond between sites $2$ and $3$.
The energy $E$ per site is thus given by
\be
E_{Dimer} = { 1 \over 2} \left[ < H_{12} > + < H_{23} > \right]
\label{Ed}
\ee
The $\phi$ dependent part of $E$ arises only from $<\hat H_{12}>$, and is
proportional to 
$ ( {\rm sin} \gamma - {\rm cos} \gamma ) {\rm cos} \phi$
with a positive coefficient.
For our region of $\gamma$, 
 $ ( {\rm sin} \gamma - {\rm cos} \gamma )  < 0 $,
hence the energy is minimized at $\phi = 0$, {\it i.e.}, 
$\zeta$ real and positive.  Henceforth we shall put
$\phi = 0$ and restrict $ 0 \le \zeta \le \infty$. 

\begin{figure}[h]
\epsfig{figure=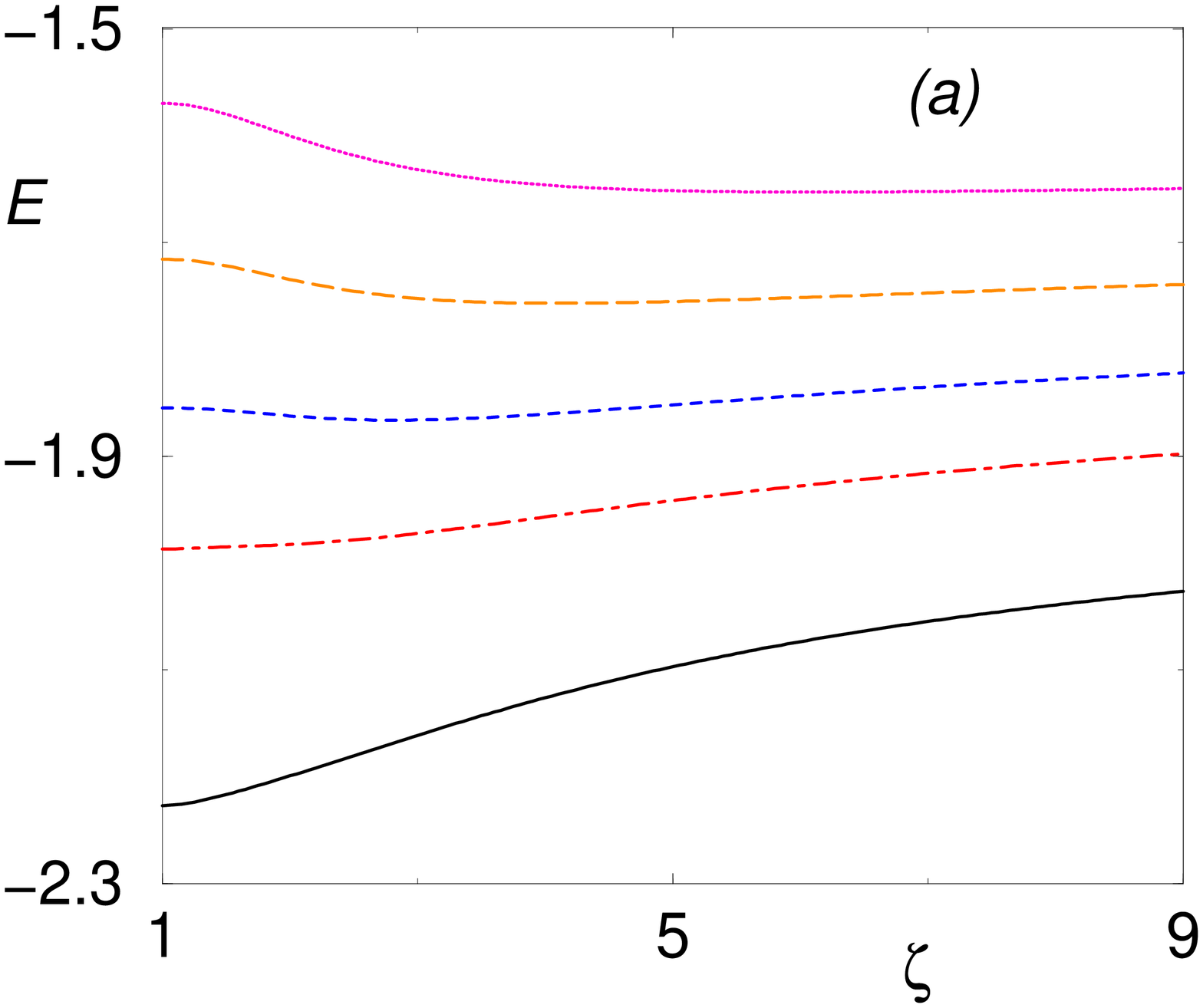,width=3.5in,angle=-90}
\epsfig{figure=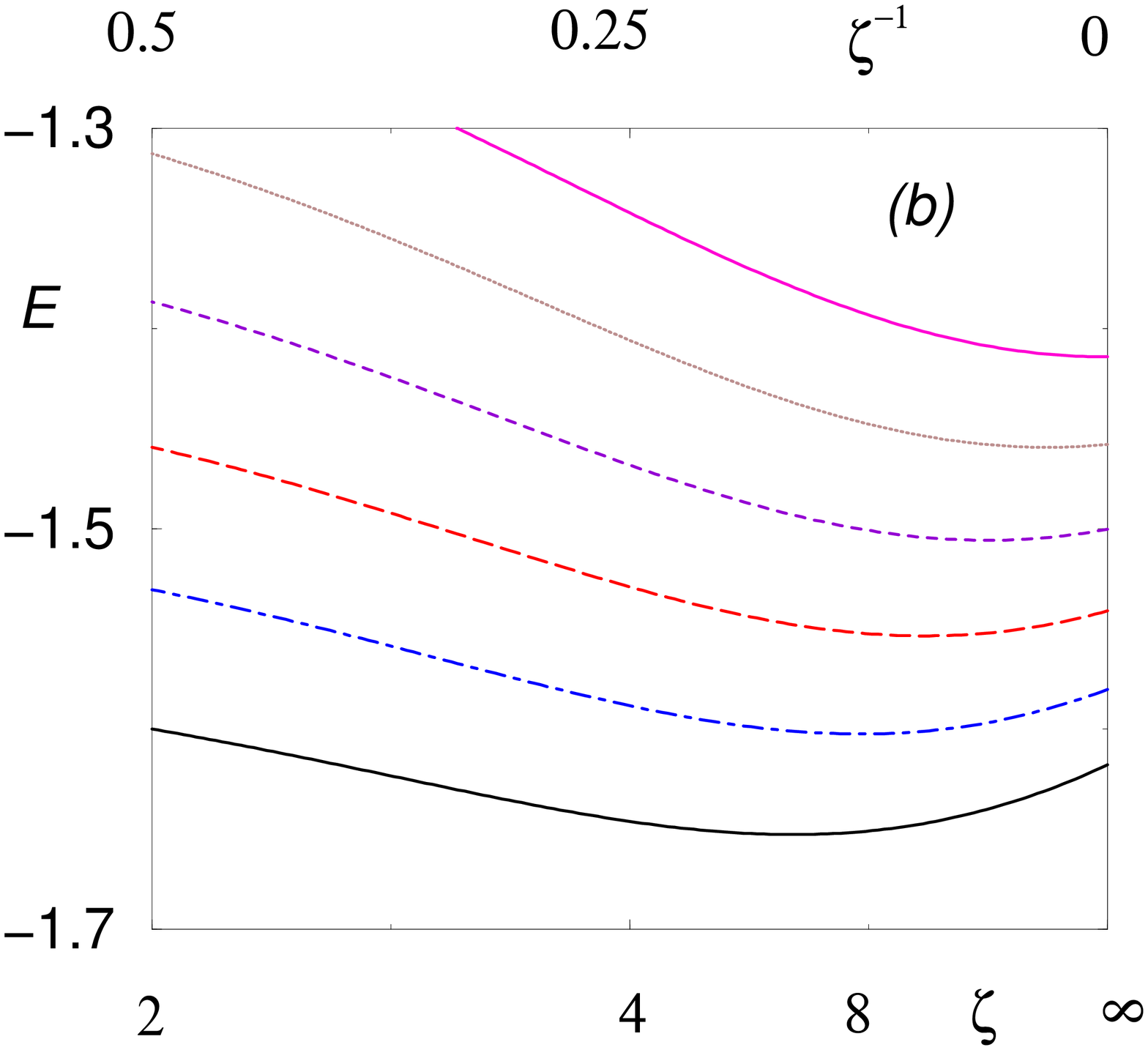,width=3.5in,angle=-90}
\vskip 0.4 cm
\begin{minipage}{0.9\textwidth}
\caption[]{ The energy $E_{Dimer}$ per site
for the ansatz wavefunction (\ref{Dimer}) (\ref{zeta})
as a function of the parameter $\zeta$.
The lines are for, from bottom to top and in
units of $\pi$, (a):
$\gamma = -0.6 $, $-0.64 $, $-0.66$,
$-0.68 $ and $-0.7$, and (b):
$\gamma = -0.7 $, $-0.71 $, $-0.72 $,
$-0.73 $, $-0.74 $ and $-0.75 $.
 }
\label{fig:E}
\end{minipage}
\end{figure}

  By expanding eq (\ref{Ed}) near $\zeta = 1$, we find
\be
E_{Dimer} = \left( - {\rm cos} \gamma + { 8 \over 3} {\rm sin} \gamma \right)
  + { 1 \over 3} \left( {\rm cos} \gamma - { 5 \over 9} {\rm sin} \gamma 
   \right) \ \left( \zeta - 1 \right) ^2  \ + \ ...
\label{E1}
\ee
Thus $\zeta = 1$ is a relative energy minimum provided 
$\gamma > \gamma_c \equiv - \pi + {\rm tan}^{-1}
{ 9 \over 5} \approx - 0.66 \pi$ defined above. 
This result is in accordance with the discussions following
eq (\ref{Psi0}).
However, we here do not interprete this as an instability of
the dimer state, but rather that a better or more generalized
ansatz is required as is done here. 
$E_{Dimer}$ as a function of $\zeta$ is plotted in Fig \ref{fig:E} (a).
  For $\gamma < \gamma_c$ we find
that the minimum energies occur at $ 1 < \zeta < \infty$.
Thus though the pair wavefunction differs from the singlet state,
the system is still dimerized.
(It can be seen easily that $ < H_{12} > \ne < H_{23} > $ 
provided $\zeta \ne \infty$.)  $\zeta$ increases without limit
when $\gamma$ decreases towards $ - { 3 \pi \over 4}$.
To examine the stability of the dimer state versus
the nematic phase, we expand eq (\ref{Ed}) as a function of 
$\eta = \zeta^{-1}$ near $\zeta = \infty$ ($\eta = 0$).  We find
\be
E_{Dimer} = 2 \ {\rm sin} \gamma + \ 2 ( {\rm sin} \gamma - {\rm cos} \gamma )
    \eta   +  ...
\label{Einf}
\ee
where the ellipsis means terms of order $\eta^2$ or higher.  Note that
the energy of the nematic phase 
is given by putting $\eta = 0$, {\it i.e.},
 $E_{Nem} = 2 {\rm sin} \gamma$.
Thus the nematic phase is never stable for
our $\gamma$ region of interest, since 
$ ( {\rm sin} \gamma - {\rm cos} \gamma ) < 0$.   The behavior
of $E_{Dimer}$ as a function of $ \zeta$ for large $\zeta$ is plotted
in Fig \ref{fig:E} (b).

Since  $ \left( {\rm sin} \gamma - {\rm cos} \gamma \right)$
changes sign at $\gamma = - { 3 \pi \over 4}$, the nematic
phase (\ref{Nem}) is more stable than the dimer state (\ref{zeta})
for $\gamma < - { 3 \pi \over 4}$.  However,  
for $- 2 \pi < \gamma < - { 3 \pi \over 4}$ the ferromagnetic
state
\be
\Psi_{Ferro} = | \ ... \ + + + \ ... >
\label{ferro}
\ee
[or any other state obtained by applying the lowering 
operator ${\bf S}_{\rm tot, -}$ arbitary 
 (limited only by twice the number of sites) 
number of times], with energy per site
$E_{ferro} = { \rm cos} \gamma + { \rm sin} \gamma$, 
becomes more stable than the nematic phase.  Thus
the nematic phase has no regime of stability for
the Hamiltonian in eq (\ref{H1D}).

The ansatz wavefunction (\ref{zeta}) does not have
definite total spin $S_{\rm tot}$ unless $\zeta = 1$.
Moreover, the wavefunction should be improved by
including spin-wave fluctuations.  However, we expect
that the projection onto definite $S_{tot}$ \cite{singlet} and correcting
the states with correlations between pairs will not change
qualitatively the picture given above.

Consider now finite $V_y$ but still $V_z \to \infty$.
We thus now have a collection of two dimensional spin-$1$ lattices.
In general the tunneling matrix elements along the $x$ and $y$
directions, hence the strength of the effective spin interactions,
can be unequal.  
The effective spin Hamiltonian for our lattice is of the same form as
eq (\ref{H}), except that we need two labels $(l_x,l_y)$
for each the lattice point, and the interactions can be between
nearest neighbor pairs along both the $x$ and $y$ directions,
with interaction constants $J_{x,y}$, $K_{x,y}$.
 With $ t_y = \delta_y^{1/2} t_x$, we have 
$J_y = \delta_y J_x$ and $K_y = \delta_y K_x$.  Without
loss in generality we can take $0 < \delta_y < 1$.
Since the excitations of
the dimer phase are gapped \cite{review,Chubukov91},
 we expect that the dimer phase
is stable at least for small $\delta_y$.  It is straight-forward
to generalize our ansatz (\ref{Dimer}) (\ref{zeta})
above to the two dimensional case,
with schematic wavefunction as shown in Fig \ref{fig:dimers} (b).
For each site, there is one "strong" bond along $x$-direction with
energy as in eq (\ref{h12}), one "weak" bond also along 
x with energy as in eq (\ref{h23}), and
two "weak" bonds along $y$ direction with total energy given by
$ 2 \delta_y$ times eq (\ref{h23}).  From this
total energy, we can see that our discussions
for one dimension is basically unaffected.  In particular
since eq (\ref{h23}) does not contribute any term linear in 
$\eta = \zeta^{-1}$, the discussions below eq. (\ref{Einf})
is qualitatively unchanged. 
Within our ansatz, the nematic phase is still always unstable
towards the dimer phase for any 
$ - { 3 \pi \over 4}  < \gamma < - { \pi \over 2}$
and $ 0 \le \delta_y < 1 $.

Under the above variational ansatz, the states are degenerate with respect
to the spatial arrangements of the pairs.  In particular,
the ansatz wavefunctions corresponding to Fig \ref{fig:dimers}
(b) and (c), with the pairs forming a 
rectangular and triangular lattice respectively,
 are completely degenerate in energy.  However, this degeneracy
will be lifted once spin-wave fluctuations are taken into account.
For $\zeta = 1$
the spin wave spectrum can be found as in ref \cite{Chubukov91}.
There are three $S=1$ modes and five $S=2$ modes, with dispersions
$\omega_1 (\vec q) = X_1 \left[ 1 - {Y_1 \nu (\vec q) \over X_1}
 \right]^{1/2}$
and 
$\omega_2 (\vec q) = X_2 \left[ 1 + {Y_2 \nu (\vec q) \over X_2}
 \right]^{1/2}$
where $X_1 \equiv {\rm cos} \gamma - 3 {\rm sin} \gamma$,
$X_2 \equiv 3 ( {\rm cos} \gamma - {\rm sin} \gamma ) $,
$Y_1 \equiv { 2 \over 3} ( 2 {\rm cos} \gamma - {\rm sin} \gamma)$
and 
$Y_2 \equiv { 2 \over 3} {\rm sin} \gamma $.
($X_1$, $X_2$, $Y_1$, $Y_2$ are all positive in our region of $\gamma$). 
Denoting the distance between sites by $ a \equiv \pi / k $,
for the rectangular lattice,
$\nu^{\rm rect} (\vec q) =  {\rm cos} (2 q_x a) + 
 2 \delta_y {\rm cos} ( q_y a)$, and
for the triangular lattice,
$\nu^{\rm tri} (\vec q) = {\rm cos} (\vec q \cdot \vec a) + 
    \delta_y  {\rm cos} ( \vec q \cdot (\vec a - \vec b) ) 
    + \delta_y {\rm cos} ( \vec q \cdot \vec b)$.
Here $\vec a \equiv  ( 2 a ) \hat x$ and
  $\vec b \equiv a ( \hat x + \hat y)$ are the lattice vectors
for our triangular lattice.
The correction to the energy from the spin-waves is given by
$\Delta E = \sum_{\vec q} \left\{
  { 3 \over 2} \left[ \omega_{1} (\vec q) -
    ( X_1 - Y_1 \nu (\vec q) \right]
+ {5 \over 2}  \left[ \omega_{2} (\vec q) -
    ( X_2 + Y_2 \nu (\vec q) \right]
\right\}$.
Evaluating this energy for small $\delta_y$, we find that
the rectangular lattice has lower energy.
Assuming that the lattice type applies for the entire region
of $\gamma$ of interest here,
we conclude that the ground state for two dimensions should
be as shown in Fig \ref{fig:dimers} (b).
Similar considerations suggest that, for three dimensions,
the lattice for ground state should be tetragonal.

The distinguishing property of the dimer state is
the doubling of the unit cell, while the spins do not
have long range order.  Thus the periodicity of the 
excitations with period $\Delta q = \pi / a =  k =  2 \pi / \lambda$
for a certain direction ($x$ above)
would be the signature of the dimer state.
These excitations can in principle be detected by
scattering.

In conclusion, we have pointed out that the trapped Bosonic
atoms provide the opportunity to realize a quantum mechanical
state predicted in theory of quantum magnetism so far
not tested experimentally.  Obtaining this state 
in the laboratory would
further widen our the play-ground for quantum many-body systems

I thank Hsiu-Hau Lin for many useful discussions.
This research was supported by the National Science Council
of Taiwan under grant number NSC91-2112-M-001-063.

\begin{thebibliography}{5}


\bibitem{Greiner02} M. Greiner {\it et al}, Nature, {\bf 415}, 39 (2002)

\bibitem{Fisher89} M. P. A. Fisher {\it et al}, Phys. Rev. B {\bf 40},
546 (1989)

\bibitem{Jaksch98} D. Jaksch {\it et al}, Phys. Rev. Lett. {\bf 81},
  3108 (1998)

\bibitem{Stenger99} J. Stenger {\it et al}, Nature, {\bf 396}, 245 (1999)

\bibitem{Ho98} T. L. Ho, Phys. Rev. Lett. {\bf 81}, 742 (1998)

\bibitem{Auerbach94} A. Auerbach, {\it Interacting Electrons and Quantum Magnetism}, Springer-Verlag, New York, 1994. 

\bibitem{com1}  The factor of $4$ is due to the product of
(a) $2$ due to two hopping processes, from left (right) to right (left) and
(b) $2$ due to identity of particles.

\bibitem{comU} $U_{0,2} > 0$ guarantee that the energy would
be higher when one well is doubly-occupied (while the other
is empty) than when each well has one single atom, as 
has been assumed here.

\bibitem{Zieger73}
H. J. Zieger and G. W. Pratt, Magnetic Interactions in Solids,
Oxford, Clarendon Press, 1973.


\bibitem{review} For review, see, e.g., I. Affleck,
 J. Phys. Condens. Matter, {\bf 1}, 3047 (1989)

\bibitem{Parkinson88} J. B. Parkinson, J. Phys. C, {\bf 21}, 3793 (1988);
  M. N. Barber and M. T. Batchelor, Phys. Rev. B {\bf 40}, 4621 (1989);
  A. Kl\"umper, J. Phys. A {\bf 23}, 809 (1990) 

\bibitem{Chubukov91} A. V. Chubukov, Phys. Rev. B {\bf 43}, 3337 (1991)

\bibitem{Fath93} G. F\'ath and J. S\'olyom, Phys. Rev. B {\bf 47}, 872 (1993)

\bibitem{Demler02} E. Demler and F. Zhou, Phys. Rev. Lett. {\bf 88},
  163001 (2002)


\bibitem{LRNO} Rigorously speaking, long range nematic order 
cannot exist in one-dimension, but we believe that taking
into account the responsible long wavelength fluctuations
would not affect much the energetics under consideration here.

\bibitem{singlet} It can be proven rigorously that the ground
state for our region of $\gamma$ has $S_{tot} = 0$.


\end{thebibliography}
\end{document}